\documentclass[10pt]{emulateapj}
\usepackage{natbib}

\shorttitle{The Quasi-Linear Premise}
\shortauthors{Howes}

\newcommand\Alfven{Alfv\'en }
\newcommand\Alfvenic{Alfv\'enic }
\newcommand{\V}[1]{\mathbf{#1}} 
 
\newcommand{\zhat}{\mbox{$\hat{\mathbf{z}}$}}

\newcommand{\figref}[1]{Figure~\ref{#1}}
\newcommand{\secref}[1]{\S\ref{#1}}
\newcommand{\eqref}[1]{equation~(\ref{#1})}

\usepackage{color}

\begin{document}


\title{The Quasilinear Premise for the Modeling of Plasma Turbulence}


\author{G.~G. Howes,  K.~G.~Klein}
\affil{Department of Physics and Astronomy, University of
Iowa, Iowa City, IA, 52242}
\and 
\author{J.~M.~TenBarge }
\affil{IREAP, University of Maryland, College Park, MD 20742}

\begin{abstract}
The quasilinear premise is a hypothesis for the modeling of plasma
turbulence in which the turbulent fluctuations are represented by a
superposition of randomly-phased linear wave modes, and energy is
transferred among these wave modes via nonlinear interactions. We
define specifically what constitutes the quasilinear premise, and
present a range of theoretical arguments in support of the relevance
of linear wave properties even in a strongly turbulent plasma. We
review evidence both in support of and in conflict with the
quasilinear premise from numerical simulations and measurements of
plasma turbulence in the solar wind. Although the question of the
validity of the quasilinear premise remains to be settled, we suggest
that the evidence largely supports the value of the quasilinear
premise in modeling plasma turbulence and that its usefulness may also
be judged by the insights gained from such an approach, with the
ultimate goal to develop the capability to predict the evolution of
any turbulent plasma system, including the spectrum of turbulent
fluctuations, their dissipation, and the resulting plasma heating.
\end{abstract}


\section{Introduction}
The presence of turbulence impacts the evolution of a wide variety of
plasma environments, from galaxy clusters to accretion disks around
compact objects, to the solar corona and solar wind, and to the
laboratory plasmas of the magnetic confinement fusion program.
Establishing a thorough understanding of plasma turbulence is a grand
challenge that has the potential to impact this wide range of research
frontiers in plasma physics, space physics, and astrophysics.
Ultimately, such efforts are aimed at developing the capability to
predict the evolution of any turbulent plasma system. The
\emph{quasilinear premise} \citep{Klein:2012} is a hypothesis for the
modeling of plasma turbulence with the potential to lead to a
quantitative, predictive theory of plasma turbulence.

The quasilinear premise states simply that \emph{some} characteristics
of the turbulent fluctuations in a magnetized plasma may be usefully
modeled by a superposition of randomly-phased, linear wave modes. The
nonlinear interactions inherent to the turbulent dynamics may be
considered to transfer energy among these linear wave
modes---therefore, the model is quasilinear. 

This premise is hotly debated at present, with significant questions
raised by the heliospheric physics community about the validity of
using the theory of linear plasma waves to analyze and interpret the
turbulent fluctuations measured in the solar wind plasma. On one hand,
a large body of work on plasma turbulence either explicitly or
implicitly assumes the relevance of some linear plasma wave
properties.\footnote{A small sample of these studies includes
  \citet{Coleman:1968,Belcher:1971,Tu:1984,Matthaeus:1990,
    Tu:1994,Verma:1995,Leamon:1998b,Quataert:1998,
    Stawicki:2001,Bale:2005,Markovskii:2006,Hamilton:2008,
    Howes:2008b,Sahraoui:2009,Schekochihin:2009,Chandran:2010a,Chen:2010,
    Podesta:2010a,Saito:2010,Rudakov:2012}. } On the other hand, the
nonlinearity inherent in turbulent interactions raises obvious
questions about the relevance of linear theory.  

In this paper, we define precisely the concepts encapsulated by the
quasilinear premise and identify the limitations of such an
approach. We outline the theoretical arguments that justify the
application of linear plasma wave theory to the study of plasma
turbulence, and review supporting and conflicting evidence from
theory, simulation, and observation.

\section{Definition of the Quasilinear Premise}
\label{sec:premise}
The quasilinear premise proposes a fundamental picture of plasma
turbulence in which the turbulent fluctuations are represented by a
superposition of randomly-phased linear wave modes, and energy is
transferred among these wave modes via nonlinear
interactions. Although this simple model cannot capture all of the
known characteristics of turbulence, it does provide a foundation upon
which a quantitatively predictive model of turbulent nonlinear energy
transfer and plasma heating may be constructed. Below we describe the
features of a model for turbulence in a magnetized plasma based on the
quasilinear premise, focusing in particular on the aspects of
turbulence that can and cannot be described by such an approach.

Several important properties of plasma turbulence are described by a
model adopting the quasilinear premise.  First, and most important, is
that the characteristic eigenfunctions of the turbulent fluctuations
(the amplitude and phase relationships among the electric, magnetic,
velocity, and density fluctuations of a single spatial Fourier mode)
are given by the linear wave eigenfunctions. Second, in a weakly
collisional plasma such as the solar wind, the fluctuations associated
with each wave mode are damped at the appropriate instantaneous
collisionless damping rate given by linear kinetic theory. Third, the
nonlinear energy transfer is described by a phenomenological
prescription derived from modern theories for anisotropic plasma
turbulence \citep{Sridhar:1994,Goldreich:1995,Boldyrev:2006,
  Howes:2008b,Schekochihin:2009,Howes:2011b}. Finally, the
distribution of the power of the turbulent fluctuations in
three-dimensional wavevector space is guided by theory and numerical
simulations of plasma turbulence, and is chosen to satisfy key
constraints, such as the observed wavenumber spectrum of magnetic
energy. Thus, the quasilinear premise is a marriage of the
quantitative linear physics of waves in a kinetic plasma with a
phenomenological prescription for the nonlinear turbulent dynamics
given by modern turbulence theories.

It is important to note that the quasilinear premise is not the same
as \emph{quasilinear theory} in plasma physics, the rigorous
application of perturbation theory to explore the long-time evolution
of weakly nonlinear systems. Rather, the quasilinear premise employs a
phenomenological prescription for the nonlinear energy transfer in
strong turbulence, rather than calculating it rigorously from first
principles. As described below in \secref{sec:theory}, the
mathematical properties of the equations that describe turbulence in a
magnetized plasma, in conjunction with a phenomenological
understanding of the properties of the turbulence, provide the
motivation for this quasilinear approach.

A model based on the quasilinear premise can be used to explore the
second-order statistical properties of plasma turbulence, such as the
energy spectrum of the turbulence or the correlations between two
different fields, for example the cross correlation between the
density and parallel magnetic field fluctuations
\citep{Howes:2012a,Klein:2012}. The \emph{synthetic spacecraft data
  method} \citep{Klein:2012} is an application of the quasilinear
premise for the interpretation of spacecraft measurements of plasma
turbulence. In this method, a synthetic plasma volume is filled with
turbulent fluctuations in the form of a distribution of
randomly-phased, linear wave modes.  This synthetic plasma volume is
then sampled along a single trajectory, generating single-point time
series of turbulent plasma and electromagnetic fluctuations that may
be analyzed with the same procedures as spacecraft measurements. A
large ensemble of such synthetic data sets has proven useful in
gaining new understanding of the compressible fluctuations in the
inertial range of solar wind turbulence \citep{Klein:2012} and of the
magnetic helicity of solar wind turbulent fluctuations as a function
of the angle between the solar wind flow and the local mean magnetic
field \citep{Klein:2014a}. In addition, one may also use the
quasilinear premise to construct a turbulent cascade model to predict
the nonlinear transfer of energy from large to small scales, enabling
predictions of the resulting energy spectra of the turbulent fields
and of the plasma heating resulting from the dissipation of the turbulent
energy via kinetic mechanisms at small scales
\citep{Howes:2008b,Podesta:2010a,Howes:2010d,Howes:2011b,Howes:2011c}.

Since the nonlinear interactions under the quasilinear premise are not
computed from the nonlinear terms in the governing equations, but
rather are given by a phenomenological prescription, the third-order
and higher order correlations observed in solar wind turbulence
\citep{Tu:1995,Bruno:2005,Hnat:2003,Hnat:2007,Kiyani:2009} cannot be investigated
using a turbulence model based on the quasilinear premise.  Such
higher order statistics depend critically on the phase relationships
between different linear wave modes, and these phase relationships are
determined by the nonlinear interactions responsible for the turbulent
cascade of energy from large to small scales. For a collection of
randomly phased linear waves, such higher order statistics will
average to zero, yielding no useful information. In other words, the
random phases between the constituent wave modes cannot capture the
intermittency and coherent structures, such as current sheets, that
are observed in plasma turbulence simulations
\citep{Wan:2012,Karimabadi:2013} and inferred from observations of
solar wind turbulence
\citep{Kiyani:2009,Osman:2012a,Osman:2012b,Wu:2013}.

In addition, inherently nonlinear fluctuations---fluctuations that
cannot be expressed as a superposition of linear
eigenfunctions---cannot be described by a model based on the
quasilinear premise. Such an inherently nonlinear fluctuation has
recently been derived in the asymptotic analytical solution of the
nonlinear interaction between counterpropagating \Alfven waves
\citep{Howes:2013a}; this mode has been shown to play an important
role in the nonlinear transfer of \Alfven wave energy to smaller
scales. However, if a phenomenological picture can be devised to
describe the distribution of such nonlinear modes, such as that
suggested by \citet{Schekochihin:2012}, it may be possible to produce
refined turbulence model by extending the quasilinear premise to
incorporate such modes.

In addition to specifying which properties of the turbulence this
approach can and cannot describe, it is also important to state which
conditions are necessary, and which are not, for the quasilinear
premise to be applicable.  First, the quasilinear premise is expected
to become valid at length scales significantly smaller than the scale
of turbulent energy injection, where the amplitude of the magnetic
field fluctuations becomes small relative to the local mean magnetic
field, $\delta B \ll B_0$.  It remains valid at all smaller scales,
including the scales at which dissipation mechanisms act to terminate
the cascade.  Second, the application of the quasilinear premise
requires both guidance from turbulence theories and constraints from
simulations and observations.  The most important element required
from turbulence theory is the prescription for the nonlinear transfer
of energy among wave modes that underlies the turbulent cascade from
large to small scales.\footnote{Note that, in principle, if the
  observed development of intermittency is taken into account in the
  phenomenological prescription chosen to describe nonlinear energy
  transfer, for example in \citet{Chandran:2014}, the resulting model
  will thereby partly account for intermittency in terms of the energy
  transfer, although the turbulent fields themselves will not be
  intermittent, so third-order statistics still cannot be explored. }
Third, the division of turbulent power among the possible linear wave
modes, and the possibly anisotropic distribution in wavevector space
of the turbulent power for each wave mode, also must be guided by
theoretical considerations as well as numerical and observational
constraints. The amplitude of the spectrum of linear modes as a
function of wavenumber can be specified such that the scaling of the
turbulent power in the model is consistent with the observed turbulent
power spectrum.  Finally, we emphasize that the fruitful investigation
of the nature of plasma turbulence using the quasilinear premise does
\emph{not} require the turbulent fluctuations derived from the linear
wave mode properties to persist for many wave periods, nor does it
require evidence of a ``linear dispersion relation'' to be apparent in
the measured turbulent power spectrum, as will discussed in more
detail in \secref{sec:theory} below.
\section{Theoretical Motivation for the Quasilinear Premise}
\label{sec:theory}
The quasilinear premise, as defined above, is motivated by the
mathematical properties of the equations that describe turbulence in a
magnetized plasma. The key concepts are most easily explained for the

simplified case of turbulence in an incompressible MHD plasma, the
minimal model containing the essential physics of plasma turbulence,
specifically the development an anisotropic \Alfvenic cascade and the
generation of current sheets at the smallest resolved scales. It is
important to note, however, that the qualitative properties
illustrated by the case of incompressible MHD turbulence are expected
to persist under more general plasma conditions, specifically carrying
over to the kinetic plasma physics describing the dissipation range of
turbulence in the weakly collisional solar wind.
\subsection{The Properties of Incompressible MHD Turbulence}
\label{sec:mhdturb}
The equations of ideal incompressible MHD can be expressed in the
symmetric form \citep{Elsasser:1950},
\begin{equation}
\frac{\partial \V{z}^{\pm}}{\partial t} 
\mp \V{v}_A \cdot \nabla \V{z}^{\pm} 
=-  \V{z}^{\mp}\cdot \nabla \V{z}^{\pm} -\nabla P/\rho_0,
\label{eq:elsasserpm}
\end{equation}
where the magnetic field is decomposed into equilibrium and
fluctuating parts $\V{B}=\V{B}_0+ \delta
\V{B} $, $\V{v}_A =\V{B}_0/\sqrt{4 \pi\rho_0}$ is the \Alfven velocity 
due to the equilibrium field $\V{B}_0=B_0 \zhat$, $P$ is total pressure (thermal
plus magnetic), $\rho_0$ is mass density, and $\V{z}^{\pm}(x,y,z,t) =
\V{u} \pm \delta \V{B}/\sqrt{4 \pi \rho_0}$ are the Els\"asser 
fields given by the sum and difference of the velocity fluctuation
$\V{u}$ and the magnetic field fluctuation $\delta \V{B}$ expressed in
velocity units \citep{Howes:2013a}. 
The Els\"asser field $\V{z}^{+}$ ($\V{z}^{-}$) represents either the
\Alfven or pseudo-\Alfven wave traveling down (up) the equilibrium magnetic
field.  The second term on the left-hand side of \eqref{eq:elsasserpm}
is the linear term representing the propagation of the Els\"asser
fields along the mean magnetic field at the \Alfven speed, the first
term on the right-hand side is the nonlinear term representing the
interaction between counterpropagating waves, and the second term on
the right-hand side is a nonlinear term that ensures incompressibility
\citep{Goldreich:1995,Howes:2013a}. 

Two properties of incompressible MHD turbulence that support the
relevance of linear wave properties to plasma turbulence are evident
upon inspection of the form of \eqref{eq:elsasserpm}.  First, as
initially recognized in the 1960s
\citep{Iroshnikov:1963,Kraichnan:1965}, if two waves are propagating
in the same direction along the mean field (yielding either
$\V{z}^{+}=0$ or $\V{z}^{-}=0$), the nonlinear interaction is zero. In
this case, an arbitrary waveform of finite amplitude provides an exact
nonlinear solution to \eqref{eq:elsasserpm} \citep{Goldreich:1995}.
Therefore, turbulent cascade of energy from large to small scales in
an incompressible MHD plasma is caused by the nonlinear interaction
between counterpropagating \Alfven waves. The physics of this
fundamental building block of astrophysical plasma turbulence has
recently been solved analytically \citep{Howes:2013a}, validated
numerically using gyrokinetic simulations \citep{Nielson:2013a}, and
verified experimentally in the laboratory
\citep{Howes:2012b,Howes:2013b,Drake:2013}.  Second, the strength of
the nonlinearity may be quantified by the magnitude of the nonlinear
term relative to the linear term, denoted by the \emph{nonlinearity
  parameter} \citep{Goldreich:1995}
\begin{equation}
\chi \equiv \frac{| (\V{z}^{\mp}\cdot \nabla) \V{z}^{\pm}|}{ |(\V{v}_A \cdot
\nabla )\V{z}^{\pm} |} \sim \frac{k_\perp v_\perp}{k_\parallel v_A}
\sim \frac{k_\perp  \delta B_\perp}{k_\parallel B_0}.
\label{eq:nlparam}
\end{equation}
 Therefore, in the limit of weak nonlinearity, $\chi \ll 1$, it is
 possible to identify a regime of weak MHD turbulence
 \citep{Sridhar:1994}, motivating a quasilinear approach using
 perturbation theory. Note that, in the case of incompressible
 hydrodynamic turbulence (Euler or Navier-Stokes), the absence of a
 linear term disallows the possibility of a perturbative approach, a
 fundamental distinction between turbulent hydrodynamic and
 magnetohydrodynamic systems.

In the absence of the nonlinear terms (setting the right-hand side of
\eqref{eq:elsasserpm} to zero), the behavior of the plasma is entirely
determined by the linear term. If the right-hand side of the equation
is considered to be an arbitrary perturbing source term, the linear
term determines the instantaneous response of the plasma to the
imposed perturbation. In the case of weak turbulence, $\chi \ll 1$
\citep{Sridhar:1994,Montgomery:1995,Ng:1996,Goldreich:1997,Ng:1997,Galtier:2000,Lithwick:2003},
the nonlinear terms on the right-hand of \eqref{eq:elsasserpm} are
indeed a small perturbation to the linear system, representing the
nonlinear transfer of energy among the linear wave modes.
Perturbation theory may be applied to the study of the turbulent
dynamics in this limit \citep{Galtier:2000,Howes:2013a,Nielson:2013a},
so the quasilinear premise is clearly valid for the case of weak
turbulence.

But the turbulent magnetic fluctuations at large scales in the solar
wind typically satisfy $\delta B/ B_0 \sim 1$, so, at the scale
of energy injection, solar wind turbulence is believed to be in a
state of strong turbulence, $\chi \sim 1$.  The theory of strong
incompressible MHD turbulence \citep{Goldreich:1995,Boldyrev:2006}
suggests that the turbulent fluctuations at small scales become
anisotropic, where the nonlinear cascade of energy generates turbulent
fluctuations with smaller scales in the perpendicular direction than
in the parallel direction, $k_\parallel \ll k_\perp$. This inherent
anisotropy of magnetized plasma turbulence has long been recognized in
laboratory plasmas \citep{Robinson:1971,Zweben:1979,Montgomery:1981}
and in the solar wind \citep{Belcher:1971}, as well as in early
numerical simulations \citep{Shebalin:1983}.  It has been conjectured
that strong turbulence in incompressible MHD plasmas maintains a state
of \emph{critical balance} between the linear timescale for \Alfven
waves and the nonlinear timescale of turbulent energy transfer
\citep{Higdon:1984a,Goldreich:1995}.  Therefore, the nonlinearity
parameter is expected to maintain a value of $\chi \sim 1$, so the
linear term $(\V{v}_A \cdot \nabla) \V{z} ^\pm $ and nonlinear term $
(\V{z}^\mp \cdot \nabla)\V{z}^\pm $ in \eqref{eq:elsasserpm} are of
the same order.  This property of strong plasma turbulence motivates
the quasilinear premise.

For strong turbulence, the condition of critical balance implies that
the energy in a particular linear wave mode may be transferred
nonlinearly to other modes on the timescale of its linear wave
period. But since the linear and nonlinear terms are of the same order
in critical balance, the linear term still contributes significantly
to the instantaneous response of the plasma, even in the presence of
strong nonlinearity. Therefore, the fluctuations in a strongly
turbulent magnetized plasma are expected to retain at least some of
the properties of the linear wave modes. In particular, for a
turbulent fluctuation with a given wavevector, the amplitude and phase
relationships between different components of that fluctuation are
likely to be related to linear eigenfunctions of the characteristic
plasma wave modes. And, the frequency response of the plasma to a
turbulent fluctuation is likely to be similar to the linear wave
frequency, at least to lowest order, the key concept used in two
recent papers used to evaluate the validity of applying the Taylor
Hypothesis \citep{Taylor:1938} to linear kinetic wave modes in the
solar wind plasma \citep{Howes:2014a} and to determine qualitatively
and quantitatively the effect of the violation of the Taylor
hypothesis on the magnetic energy frequency spectrum measured \emph{in
  situ} by spacecraft \citep{Klein:2014b}.

It is important to address here a common misconception that plasma
turbulence with small amplitude fluctuations, $\delta B/ B_0 \ll 1$,
is necessarily weak turbulence. At small scales in the inertial range,
and throughout the dissipation range, turbulent fluctuations of the
magnetic field satisfy the condition $\delta B/ B_0 \ll 1$.  The
small amplitude of the fluctuations has lead many researchers to
conclude that the turbulent interaction must necessarily be
weak. However, as can be seen from the inspection of
\eqref{eq:nlparam}, the ratio of the nonlinear to the linear term in
the equation is given by $\chi \sim (k_\perp \delta
B_\perp)/(k_\parallel B_0)$.  Therefore, if the turbulent fluctuations
are sufficiently anisotropic, $k_\perp/k_\parallel \gg 1$, then the
nonlinearity parameter may easily yield the condition of strong
turbulence, $\chi \sim 1$, even when fluctuations have a small
amplitude, $\delta B_\perp/ B_0 \ll 1$. Therefore, without knowledge
of the anisotropy of the turbulent fluctuations, it is incorrect to
conclude that turbulence is weak simply because the fluctuation
amplitudes are small.

Given that the idea of critical balance \citep{Goldreich:1995} in
strong plasma turbulence is essentially a quasilinear concept---that
the frequency of the nonlinear energy transfer remains of the same
order as the \emph{linear} wave frequency---evidence in support of
critical balance indirectly supports the quasilinear premise.
Although the concept of critical balance was initially applied only to
the case of MHD turbulence, the theory has been extended to the
dispersive wave regime of the dissipation range
\citep{Biskamp:1999,Cho:2004,Krishan:2004,Shaikh:2005,
  Galtier:2006,Howes:2008b,Schekochihin:2009}.  Evidence in support of
or in conflict with the conjecture of critical balance from numerical
simulations and solar wind observations is reviewed in
\secref{sec:num} and \secref{sec:obs} below.

\subsection{Analogy: Critically Damped Harmonic Oscillator}
\label{app:sho}
The simple physical example of a damped simple harmonic oscillator
illustrates the concept, central to the quasilinear premise, that the
linear terms of an equation contribute significantly to the evolution
of the system, even in the absence of oscillatory (or wave-like)
behavior. The equation of evolution for a damped harmonic oscillator is 
given by 
\begin{equation}
\frac{d^2x}{dt^2} + \omega_0^2 x + \nu \frac{dx}{dt} = 0,
\label{eq:dsho}
\end{equation}
where $\omega_0$ is the undamped frequency of the harmonic oscillator
and $\nu$ is a frictional damping coefficient. The dimensionless
parameter that characterizes the strength of the damping relative to
the linear restoring term responsible for oscillatory behavior is
$\zeta= \nu/(2\omega_0)$. This term is analogous to the nonlinearity
parameter $\chi$ in plasma turbulence which characterizes the strength
of the nonlinear to the linear term, given by \eqref{eq:nlparam}.  The
system is underdamped for $\zeta < 1$, critically damped for $\zeta =
1$, and overdamped for $\zeta > 1$.

\begin{figure*}[t]
\vspace*{2mm}
\begin{center}
\hbox to \hsize{\resizebox{18.cm}{!}{
\includegraphics*{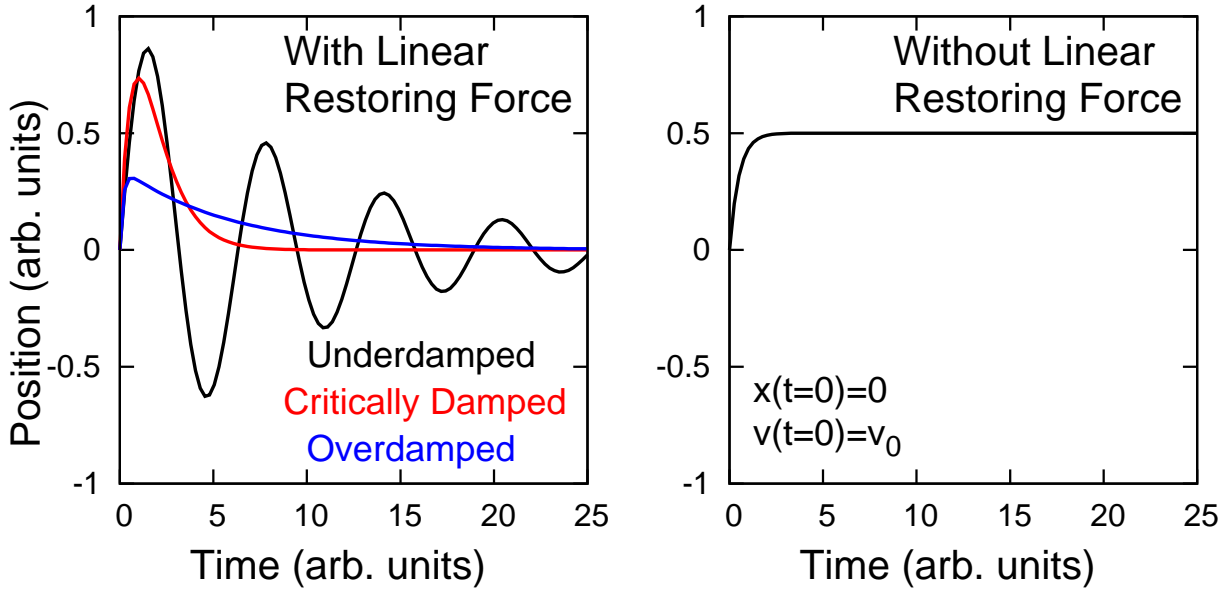}}}
\end{center}
\caption{(a)~The position $x$ vs.~time $t$ for a damped simple
  harmonic oscillator, given by \eqref{eq:dsho}, for an underdamped
  $\zeta < 1$ case (dashed), a critically damped $\zeta = 1$ case
  (solid), and an overdamped $\zeta > 1$ case (dotted). (b)~For a
  system in which restoring force is eliminated ($\omega_0=0$), the
  position $x$ vs.~time $t$ is qualitatively different.}
\label{fig:dsho}
\end{figure*}

Specifying the initial conditions at $t=0$ to be $x=0$ and
$dx/dt=v_0$, we consider the evolution of the system in time, as shown
in panel (a) of \figref{fig:dsho}. In the underdamped $\zeta < 1$ case
(dashed), the oscillatory behavior of the system is evident. For
either critically damped $\zeta = 1$ case (solid) or overdamped $\zeta
> 1$ case (dotted), the system does not demonstrate oscillatory
behavior.  In either of these non-oscillatory cases, however, the
linear restoring term still plays an important role in governing the
evolution of the system.

If the restoring term is eliminated by setting $\omega_0=0$ in
\eqref{eq:dsho}, the initial velocity (not shown) monotonically
decreases to zero and the position asymptotes to a value
$x_\infty=v_0/\nu$, as shown in panel (b) of \figref{fig:dsho}. But
when the linear restoring term is present, the evolution is
qualitatively different, as seen by comparing the left and right
panels of \figref{fig:dsho}.  For the critically damped case (right
panel, solid), the position reaches maximum value $x<x_\infty$ before
returning monotonically to zero, while the velocity (not shown)
decreases from the initial velocity $v_0$ to a minimum $v<0$, before
returning monotonically to zero.  Therefore, even though the damped
simple harmonic oscillator for $\zeta \ge 1$ exhibits no oscillatory
behavior, the presence of the linear restoring term still
significantly influences the evolution of the system.

In a turbulent magnetized plasma, the balance is between the nonlinear
and linear response terms, rather than linear damping and linear
restoring terms, but the lesson is analogous: even in a strongly
turbulent plasma in which the nonlinear term is of the same order as
the linear term, the linear terms continue to play a significant role
in governing the turbulent dynamics. The phenomenon of critical
damping in this harmonic oscillator system is analogous to the concept
of critical balance in plasma turbulence. It is important to point out
that this analogy does \emph{not} carry over to the case of
hydrodynamic turbulence---the absence of a linear wave term in the
governing equations precludes the identification of an equivalent
critical parameter. This highlights a fundamental difference in the
nature of hydrodynamic and plasma turbulence.
\subsection{Is Evidence of a Linear Dispersion Relation Necessary?}
\label{sec:disp}

The concept central to the quasilinear premise that the turbulence
consists of a broadband spectrum of \Alfven waves is frequently
criticized with the argument that, if waves are indeed present in the
turbulence, one should be able to see clear evidence of the linear
dispersion relation (meaning a plot of $\omega$ vs.~$k$) in the
analysis of turbulence simulations. We argue here that there are two
good reasons why one should not necessarily be able to see clearly
such a dispersion relation. The first reason is related to the
semantic difference between an \Alfven wave and an \Alfvenic
fluctuation.  The second reason is that changes in the amplitude or
phase of a wave with a constant frequency will significantly broaden
the frequency content determined by a Fourier transform of a time
series.

In the weak turbulence limit, $\chi \ll 1$, it requires nonlinear
interactions with many counterpropagating \Alfven waves before the
bulk of the energy of a given \Alfven wave is cascaded to a smaller
scale \citep{Sridhar:1994,Howes:2013a}. In this limit, therefore, it
is likely that the signature of wave-like motions may be relatively
easy to observe. But for a case of strong turbulence, in which the
nonlinearity parameter $\chi \sim 1$, the energy of an \Alfven wave is
completely cascaded to smaller scale through a collision with a single
counterpropagating \Alfven wave, so that the timescale for nonlinear
energy transfer is the same order as the linear wave period. In this
case, if the interacting turbulent fluctuations persist for only a
single wave period, does it make sense to refer to the interacting
fluctuations as waves?  In our view, this point is largely semantic,
for the following reasons.

In the simplified context of incompressible MHD, any finite amplitude
fluctuation with $\V{z}^+\ne 0$ and $\V{z}^-=0$ (or with $\V{z}^+= 0$
and $\V{z}^-\ne 0$ ), is an exact nonlinear solution of the equations
of evolution, corresponding to an arbitrary waveform propagating
unchanged in one direction along the mean magnetic field at the
\Alfven velocity. This solution is clearly a finite amplitude \Alfven
wave, but it need not oscillate sinusoidally as one typically
envisions when discussing a wave-like motion.  A Fourier transform of
the time series measured at a fixed Eulerian point as the wave passes
by may yield a broad frequency response; for example, an isolated
``wavepacket'' consisting of a single parallel wavelength
$\lambda_\parallel$ of a perfectly sinusoidal signal will indeed
return a frequency response that is significantly broadened about the
``linear'' frequency $\omega=2 \pi v_A/\lambda_\parallel$.  Some may
prefer the more general term ``\Alfvenic fluctuation'' in place of
``\Alfven wave'' in this case, be the difference is merely
semantic. That an \Alfven wave may not be purely sinusoidal, and thus
not have a well defined frequency $\omega$, does not change the nature
of the fluctuation, which is determined by the linear response of the
plasma to an applied perturbation, as dictated by the linear term in
the equation of motion \eqref{eq:elsasserpm}. It is in this sense that
we speak of turbulence supported by \Alfven waves.  Therefore, we
hereby clarify that our use of the term \Alfven wave pertains to any
fluctuation whose linear response is the same as that of an \Alfven
wave---thus, we refer to any \Alfven fluctuation as an \Alfven wave.

\begin{figure}[t]
\resizebox{8.3cm}{!}{
\includegraphics*{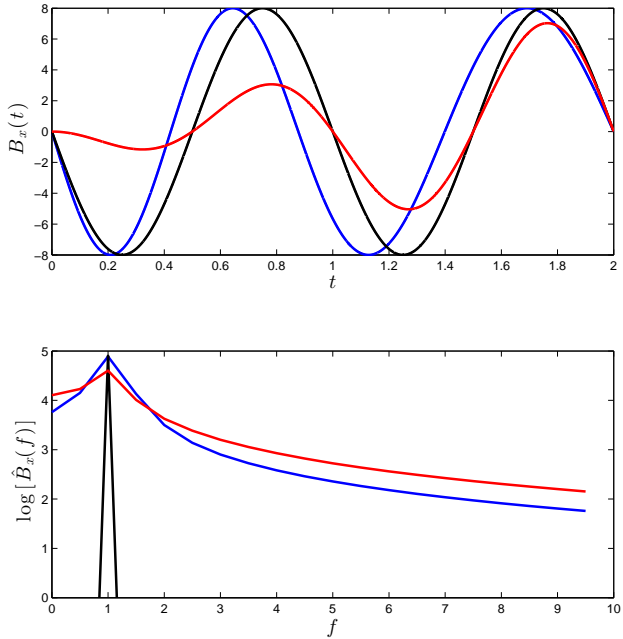}}
\caption{(a) Pure sinusoidal signal (black), sinusoidal signal with linearly increasing amplitude (red), and sinusoidal signal with shifting phase (blue) as a function of time. (b) Fourier transform of the signals in panel (a), showing the significantly broadened response for changes in amplitude or phase.}
\label{fig:phase_amp}
\end{figure}

The second reason that evidence of a linear dispersion may not be
apparent in measurements or simulations of strong plasma turbulence,
even for turbulence consisting precisely of a sum of linear wave
modes, is that changes in the amplitude or phase of a wave will
broaden the frequency response substantially, as shown in
\figref{fig:phase_amp}. Here we plot the frequency retrieved from a
Fourier transform of the time series for a pure sinusoidal signal
(black), a sinusoidal signal with linearly increasing amplitude (red),
and a sinusoidal signal with shifting phase (blue).  It is clear that
the frequency response when the amplitude or phase of a mode changes,
one obtains a very broadened response in frequency, even when the
basic signal has a well-defined frequency. Therefore, it is likely to
be very difficult to recover a clean $\omega$ vs. $k$ dispersion
relation, even for turbulence simulation data.

To make matters more complicated, for spacecraft measurements, the
presence of a broad spectrum of \Alfven wave power in
three-dimensional wavevector space further complicates the
interpretation of a time series of measurements at a fixed point in
space in terms of the frequency of the underlying fluctuations.

It is worthwhile to raise one final point regarding the frequency
spectrum of the fluctuations in plasma turbulence. One may view the
dynamical evolution of plasma according to \eqref{eq:elsasserpm} as
the linear response of the plasma (given by the left-hand side of the
equation) to perturbations generated by the nonlinear terms (given by
the right-hand side of the equation). For a particular wavevector
$\V{k}$, the characteristic frequency of the linear plasma response is
given by $\omega_0=k_\parallel v_A$, where the component of the
wavevector parallel to the mean magnetic field is $k_\parallel = \V{k}
\cdot \V{B}_0/B_0$. As in the case of a driven, damped single harmonic
oscillator, one may obtain a response at a frequency $\omega \ne
\omega_0$ only by sustained driving at another frequency.  But since
the nonlinear terms perturbing a given wavevector mode are dominated
by the interactions between nearby wavevectors, the frequency driving
a particular spatial Fourier mode in plasma turbulence is likely to be
very near the linear frequency of the mode being driven.  The power
generated by the nonlinear interaction between two counterpropagating
\Alfven waves cannot have a frequency higher than the sum of the
frequency of those two modes \citep{Howes:2013a}, so the turbulent
power will not contain frequencies significantly higher than the
characteristic linear \Alfven wave frequency for a particular
wavevector.  In addition, there would be a significant impedance
mismatch if the plasma is driven at a frequency well above the linear
\Alfven wave frequency, as seen in simulations
\citep{Parashar:2011,TenBarge:2012a,TenBarge:2014a}.

\subsection{Collisionless Damping of Turbulent Fluctuations}
\label{sec:damp}

An important component of the quasilinear premise that enables the
construction of predictive models of the turbulent dissipation and
resulting plasma heating is that the turbulent fluctuations associated
with each wave mode are damped at the appropriate instantaneous
collisionless damping rate given by linear kinetic theory. The
applicability of these linear kinetic damping rates, of course,
depends on the central concept of the quasilinear premise that the
turbulent fluctuations have the properties of the linear waves
supported by the weakly collisional plasma.  It is important to note
that the terms in the kinetic equation for weakly collisional plasmas
that are responsible for collisionless wave-particle interactions with
the equilibrium distribution function are linear, and therefore the
same arguments for the relevance linear wave modes in a turbulent
plasma also apply to the linear collisionless damping.  Here we
discuss specifically two important issues regarding the collisionless
dissipation: (i) possible nonlinear saturation or inhibition of the
collisionless damping and (ii) the spatial distribution of damping
via wave-particle interactions.

There are two possible ways that the linear collisionless damping rate
may be altered: (i) quasilinear or nonlinear evolution of the
equilibrium distribution function may flatten the slope at the
resonant velocity for a particular wave, thus reducing or entirely
inhibiting collisionless damping of that wave \citep{Rudakov:2011,Rudakov:2012};
and (ii) forcing by strong turbulence in a stochastic manner may break
the coherence between a particular wave mode and a particle necessary
for collisionless damping to extract significant energy from the wave
\citep{Plunk:2013,Kanekar:2014b}.

It has been suggested that electron Landau damping is the dominant
collisionless mechanism for the damping of fluctuations in the
dissipation range of solar wind turbulence
\citep{Leamon:1998b,Leamon:1999,Leamon:2000,Howes:2008b,Schekochihin:2009,Howes:2009b,TenBarge:2013a,TenBarge:2013b}.
In this regime, it has been proposed that the turbulent fluctuations
are kinetic \Alfven waves, a hypothesis now strongly supported by a
number of numerical
\citep{Howes:2008a,Howes:2011a,Boldyrev:2012b,TenBarge:2013a,TenBarge:2013b}
and observational studies
\citep{Belcher:1971,Harmon:1989,Leamon:1998a,Hollweg:1999,Bale:2005,Smith:2006,Hamilton:2008,Howes:2008b,Chandran:2009b,Gary:2009,Howes:2010a,Sahraoui:2010b,He:2011,Podesta:2011a,Podesta:2012,Salem:2012,TenBarge:2012b,Chen:2013b,Klein:2014a}
of solar wind turbulence.  See \citet{Podesta:2013} for a recent
review of evidence for kinetic \Alfven waves in the solar wind.
Flattening of the distribution function, and the associated reduction
in collisionless damping rates, due to quasilinear evolution of the
equilibrium is a well-understood process in plasma physics. The key
question in the context of the turbulent solar wind plasma is whether
this mechanism operates effectively to quench collisionless damping,
as recently suggested \citep{Rudakov:2011,Rudakov:2012}.

An argument against this quasilinear quenching of the Landau damping
of kinetic \Alfven waves arises from the variation in the resonant
velocity in the electron distribution function.  Since the kinetic
\Alfven wave is a dispersive wave mode, its phase velocity (which is
nearly parallel to the magnetic field direction) increases linearly
with $k_\perp$ in the dissipation range at $k_\perp\rho_i \gtrsim 1$.
Therefore, the parallel velocity of electrons that are resonant with
the turbulent kinetic \Alfven waves\footnote{Here we assume that the
  ion and electron temperatures are roughly equal, $T_i \sim T_e$.}
increases from a velocity $v_\parallel \ll v_{te}$ at $k_\perp \rho_i
\sim 1$ to $v_\parallel \sim v_{te}$ at $k_\perp \rho_e \sim 1$.
Therefore, without flattening the electron velocity distribution over
the entire range $v_\parallel \le v_{te}$, it seems unlikely that a
quasilinear flattening of the distribution function could completely
suppress Landau damping for $k_\perp \rho_e \le 1$.

A second possible way to suppress collisionless wave-particle
interacts is to interfere with the coherence time between
electromagnetic waves and particles so that the particles are unable
to experience a net gain of energy.  For example, in strongly
collisional plasmas, such as MHD plasmas, electromagnetic waves are
undamped by collisionless wave-particle interactions because frequent
collisions prevent a single particle from maintaining a sufficiently
long coherence time to interact resonantly with the waves. A recent
study has shown that, when the linearized Vlasov equation is perturbed
by a stationary random force, the effective Landau damping rate can be
significantly reduced under appropriate conditions
\citep{Plunk:2013}. Work continues to understand the effect of Landau
damping in a turbulent setting \citep{Kanekar:2014b}. Nonlinear
kinetic simulations of plasma turbulence will hopefully be able to
quantify any suppression of the linear kinetic damping rate.

Another issue associated with the collisionless damping of plasma
turbulence is the spatial distribution of the plasma heating resulting
from the dissipation of the turbulent energy. Based on the
decomposition of turbulent fluctuations into component plane waves, it
has frequently been asserted that Landau damping must lead to
spatially uniform heating. Recently, there has been significant work
investigating non-uniform distribution of heating in the solar wind
plasma, especially focusing on enhanced heating in the neighborhood of
current sheets \citep{Osman:2011,Osman:2012a,Osman:2012b}, where it has been 
suggested that heating associated with current sheets cannot be a
consquence of Landau damping.  However, the assertion that Landau
damping must lead to spatially uniform heating is simply incorrect. 

Consider the case of a current sheet with a strong guide magnetic
field, meaning that the current sheet is associated with an abrupt
rotation of the angle of the magnetic field.  The spatially localized
structure of the current sheet necessarily implies significant energy
in small wavenumber components. If one views the situation in
wavevector space, each of the Fourier components Landau damps at its
linear kinetic damping rate.  But, the spatial distribution of this
heating will not be uniform, but instead will be localized in the
vicinity of the current sheet.  This argument has recently been made
to resolve the apparent contradiction in recent findings from
gyrokinetic simulations of turbulence at the electron scales where a
clear correlation between electron heating rate and the presence of
current sheets is observed, but the heating as a function of
wavenumber is completely explained, with no fitting parameters, by
Landau damping of each spatial Fourier mode at it linear kinetic
damping rate \citep{TenBarge:2013a}.

\section{Supporting Evidence from Numerical Simulations}
\label{sec:num}

Besides the feasibility arguments for the validity of the quasilinear
premise outlined above, we give no \emph{a priori} proof for its
validity in strongly nonlinear plasma turbulence. Nonlinear
simulations of plasma turbulence provide a powerful and direct avenue
for testing the validity of the premise.  In this section, we present
evidence from numerical simulations that support the relevance of
linear plasma wave properties in a strongly turbulent plasma,
including evidence of the relevance of linear wave eigenfunctions,
linear collisionless damping rates, and linear frequency response as
well as support for critical balance.

First, a gyrokinetic simulation of the transition from the inertial to
the dissipation range found that the perpendicular magnetic, parallel
magnetic, and perpendicular electric field energy spectra are well fit
by a turbulent cascade model \citep{Howes:2008b} that assumes the
relationships between these energy spectra are determined solely by
the linear kinetic eigenfunctions of the Alfv\'en and kinetic \Alfven
waves \citep{Howes:2008a}.  Second, a gyrokinetic simulation spanning
the entire dissipation range from the ion to the electron Larmor
radius was able to predict the parallel magnetic and perpendicular
electric field energy spectra observed in the simulation by using the
perpendicular magnetic energy spectrum and assuming that the
relationships between the turbulent field components are described by
the linear eigenfunction of the kinetic \Alfven wave
\citep{Howes:2011a}. These energy spectra were also well fit by a
refined cascade model that again assumed the applicability of linear
eigenfunctions and kinetic damping rates \citep{Howes:2011b}. Finally,
an analysis of a simulation of kinetic \Alfven wave turbulence finds
that the measured electron heating as a function of perpendicular
wavenumber is well estimated, with no fitting parameters, by assuming
all dissipation is provided by collisionless wave particle
interactions at the linear Landau damping rate \citep{TenBarge:2013a}.

Another proposed technique for evaluating the importance of linear
wave modes is to compute the frequency spectrum for particular Fourier
modes or the Eulerian frequency spectrum at a particular point in
space from numerical simulations \citep{Dmitruk:2009}. As discussed
above in \secref{sec:disp}, except for the case of weak turbulence,
however, it has not been established that one should indeed expect to
see evidence of a ``linear dispersion relation'' in the frequency
spectra of strong, driven plasma turbulence since an individual wave
mode is not expected to persist for more than a single wave period. A
study of driven turbulence in 3D incompressible MHD simulations found
little evidence that linear waves play a significant role in MHD
turbulence \citep{Dmitruk:2009}, but it has been pointed out that the
method of driving used in the simulations may have had a dominant
impact on the measured frequency spectrum
\citep{TenBarge:2012a,TenBarge:2014a}. A particle-in-cell (PIC)
simulation of decaying 2D magnetosonic turbulence over the
$k_\parallel$-$k_\perp$ plane found a cascade consistent with the
properties of fast magnetosonic waves, and that little energy appeared
to be nonlinearly transferred to the slow magnetosonic or ion
Bernstein waves \citep{Svidzinski:2009}. Two studies of driven, 2D
hybrid kinetic ion and fluid electron simulations of turbulence over
the perpendicular plane found a low level of wave activity, with the
dynamics dominated by nonlinear activity
\citep{Parashar:2010,Parashar:2011}. The 2D geometry of these hybrid
simulations, however, admits only linear wave modes with
$k_\parallel=0$, and therefore these simulations cannot support
\Alfven waves, kinetic \Alfven waves, or whistler waves, calling into
question their relevance to the study of solar wind
turbulence. Finally, 3D decaying simulations of turbulence at the
high-frequency end of the inertial range using both the Hall MHD and
Landau fluid theory \citep{Passot:2007} find that the peak of the
turbulent magnetic and kinetic energy frequency spectra for particular
Fourier modes show excellent agreement with the linear wave mode
frequencies for the fast, Alfv\'en, and slow modes
\citep{Hunana:2011}. 
 This collection of apparently contradictory
findings suggests that this line of investigation of the relevance of
linear wave properties in numerical simulations of turbulence will
remain an active area of research.

Given that the idea of critical balance \citep{Goldreich:1995} in
strong plasma turbulence is essentially a quasilinear concept---that
the frequency of the nonlinear energy transfer remains of the same
order as the \emph{linear} wave frequency---numerical evidence in
support of critical balance indirectly supports the quasilinear
premise.  Although the concept of critical balance was initially
applied only to the case of MHD turbulence, the theory has been
extended to the dispersive wave regime of the dissipation range
\citep{Biskamp:1999,Cho:2004,Krishan:2004,Shaikh:2005,
  Galtier:2006,Howes:2008b,Schekochihin:2009,Howes:2011b}. In
numerical simulations, critical balance is generally tested by a a
measure of the turbulent power on the $k_\perp$-$k_\parallel$ plane in
wavevector space in numerical simulations
\citep{Cho:2000,Maron:2001,Cho:2004,Grappin:2010}.  The results of
these numerous studies are contradictory, with some claiming to
support critical balance, and others, to refute it.  The lines of
conflict, however, coincide with the method used to determine the
direction of the magnetic field: studies using a \emph{local} mean
magnetic field \citep{Cho:2000,Maron:2001,Cho:2004} are consistent
with the predictions of critical balance, while studies employing a
global magnetic field \citep{Grappin:2010} are inconsistent with
critical balance. It appears that, as long as the direction of the
magnetic field is determined locally, there exists significant
evidence in support of critical balance.

A recently proposed alternative method for testing critical balance
eliminates the need to define the direction of the magnetic field by
noting that the linear wave frequency for \Alfvenic plasma waves is
proportional to the parallel wavenumber, $\omega \propto k_\parallel$
\citep{TenBarge:2012a}. In this case, one may compute the distribution
of turbulent power on the $\omega$-$k_\perp$ plane and look for the
predicted scaling $\omega \propto k_\parallel \propto k_\perp^\alpha$,
where $\alpha=2/3$ \citep{Goldreich:1995} or $\alpha=1/2$
\citep{Boldyrev:2006} for the MHD \Alfven wave cascade and
$\alpha=1/3$ for the kinetic \Alfven wave cascade
\citep{Cho:2004,Howes:2008b,Schekochihin:2009}. Gyrokinetic
simulations of driven, 3D kinetic \Alfven wave turbulence support the
predicted critical balance scaling of $\alpha=1/3$ in this regime
\citep{TenBarge:2012a,TenBarge:2013b}.

Finally, a recent study of ion temperature gradient driven turbulence
in gyrokinetic simulations of magnetic confinement fusion plasmas has found support
for the turbulent fluctuations satisfying critical balance \citep{Barnes:2011}.

\section{Supporting Evidence from Solar Wind Observations}
\label{sec:obs}

Spacecraft measurements in the turbulent wind plasma have been used to
evaluate the characteristic nature of the turbulent fluctuations,
seeking agreement with the typical frequencies of \Alfven and kinetic
\Alfven wave modes using a k-filtering analysis of multi-spacecraft
data and searching for evidence of critical balance through the
spectral indices measured at different angles with respect to the
direction of magnetic field. In addition, laboratory experiments of
plasma turbulence have begun to uncover evidence suggesting the turbulence
indeed satisfies critical balance.

Multi-spacecraft observations of solar wind turbulence may also be
used to determine the fluctuation frequency in the solar wind plasma
frame for each spatial Fourier mode
\citep{Sahraoui:2010b,Narita:2011,Roberts:2013} .  The first
$k$-filtering analysis of \emph{Cluster} multi-spacecraft data in the
unperturbed solar wind showed that the turbulent fluctuations in the
inertial and transition range, $0.04 \le k_\perp \rho_i \le 2$, are
consistent with the dispersion relation of the Alfv\'en/kinetic
\Alfven wave branch, and are inconsistent with the fast/whistler
branch \citep{Sahraoui:2010b}.  A subsequent study that performed a
similar $k$-filtering analysis of \emph{Cluster} multi-spacecraft data
obtained contradictory results, finding little agreement with any
particular linear wave mode \citep{Narita:2011}. A number of issues,
however, cast serious doubts on the validity of the latter study: the
error in the plasma-frame frequency is not estimated, the results of
this study at large scales are inconsistent with the observations that
demonstrate the largely incompressible, \Alfvenic nature of large
scale fluctuations in the solar wind
\citep{Belcher:1971,Tu:1995,Bruno:2005}, inspection of the CIS and
PEACE data show that all of the intervals studied suffer either
electron or ion foreshock contamination, and the spacecraft
configuration for each of these intervals shows significant levels of
planarity and elongation, indicating a poor tetrahedron. A subsequent
k-filtering study found further evidence in support of low frequency
kinetic \Alfven waves in \emph{Cluster} multi-spacecraft data
\citep{Roberts:2013}.  It is clear that new multi-spacecraft studies
will provide valuable guidance in assessing the relevance of linear
wave modes in the solar wind plasma.

In measurements of solar wind turbulence, one can test the idea of
critical balance \citep{Goldreich:1995} by measuring the 1D magnetic
energy spectrum as a function of the angle $\theta_{VB}$ with respect
to the magnetic field
\citep{Horbury:2008,Podesta:2009a,Tessein:2009,Chen:2010,Wicks:2010,Forman:2011}.
As in the case for numerical simulations, there exist contradictory
findings, with support for critical balance found when the magnetic
field direction is determined locally
\citep{Horbury:2008,Podesta:2009a,Chen:2010,Wicks:2010,Forman:2011}
and conflict with critical balance when the study employs a global
magnetic field to determine the parallel direction
\citep{Tessein:2009}.

Finally, recent experimental measurements of strongly magnetized plasma
turbulence driven by the ion temperature gradient in the Mega Amp
Spherical Tokamak (MAST) have been found to be consistent with the predictions
of critical balance \citep{Ghim:2013}.

\section{Discussion}
\label{sec:discuss}
The question of the validity of the quasilinear premise---that linear
wave properties are relevant to the dissipation range of solar wind
turbulence---clearly remains to be settled. On balance, however, the
bulk of the evidence appears to argue in favor of its validity. The
utility of the quasilinear premise for the study of plasma turbulence,
however, may also be judged \emph{a posteriori} by the insights gained
from such an approach. Below we briefly discuss a number of
applications of the quasilinear premise to the study of plasma
turbulence.

The \emph{synthetic spacecraft data method} \citep{Klein:2012} is a
technique for producing artificial plasma turbulence measurements that
can be directly compared to \emph{in situ} measurements of turbulence
in the solar wind.  The turbulence in the synthetic plasma volume is
constructed assuming the quasilinear premise.  Such an approach has
proven to be very successful, discovering that the compressible
fluctuations in the inertial range of solar wind turbulence are
anisotropically distributed slow waves \citep{Howes:2012a,Klein:2012},
and illuminating the nature of the fluctuations responsible for the
observed magnetic helicity signature of turbulent fluctuations as a
function of the angle between the magnetic field and the solar wind
flow \citep{Klein:2014a}.

Additionally, based on the fundamental concepts embodied by the
quasilinear premise, a number of simple models for the turbulent
cascade of energy in weakly collisional plasma turbulence have been
devised \citep{Howes:2008b,Podesta:2010a,Howes:2011b}, with the
ability to model the dissipation of the turbulent cascade based on
kinetic damping mechanisms \citep{TenBarge:2012b}. In addition to
successfully modeling the energy spectra in nonlinear kinetic
simulations of turbulence \citep{Howes:2008a,Howes:2011b}, such
cascade models can be applied to develop a predictive capability for
the plasma heating resulting from the dissipation of turbulence in
weakly collisional space and astrophysical plasmas
\citep{Howes:2010d,Howes:2011c}.

Finally, the characteristic eigenfunctions of the linear kinetic
plasma waves can be exploited in an attempt to identify the
characteristic nature of the turbulent solar wind fluctuations
\citep{Sahraoui:2009,Howes:2010a,Gary:2010,Podesta:2011a,He:2011,Salem:2012,Smith:2012,Chen:2013a,Chen:2013b}.

There are also a couple of other lines of argument that support the
relevance of linear wave properties to the strongly turbulent solar
wind plasma. First, an important feature of solar wind turbulence is
the observation that linear kinetic temperature anisotropy
instabilities appear to constrain the limits of the temperature
anisotropy $T_\perp/T_\parallel$ of ions in the solar wind
\citep{Kasper:2002,Hellinger:2006,Bale:2009}. Second, from 
fundamental theorems about dimensional analysis, \emph{i.e.} the Pi Theorem, the
dimensionless parameters upon which the linear theory depends will
necessarily also be parameters upon which the nonlinear theory depends
(since linearization involves dropping terms, never adding), although
the nonlinear theory undoubtedly depends on additional dimensionless
parameters.

\section{Conclusion}
\label{sec:conc}
The quasilinear premise is a hypothesis for the modeling of plasma
turbulence in which the turbulent fluctuations are represented by a
superposition of randomly-phased linear wave modes, and energy is
transferred among these wave modes via nonlinear interactions.
Although a large body of work on plasma turbulence either explicitly
or implicitly assumes the relevance of some linear plasma wave
properties, the nonlinearity inherent in turbulent interactions raises
obvious questions about the relevance of linear theory. This papers
attempts to present a broad range of theoretical, numerical, and
observational evidence in the attempt to evaluate the validity of the
quasilinear premise.

After defining the quasilinear premise precisely, we highlight the the
aspects of turbulence that can and cannot be described by such an
approach: turbulent fluctuation properties such as the eigenfunction,
frequency, and collisionless damping rate as well as second-order
statistics such as the energy spectra or magnetic helicity of the
turbulence can be described by a model of turbulence based on the
quasilinear premise; third-order and higher order statistics, such as
intermittency and coherent structures, such as current sheets, cannot
be investigated using such an approach.

We present a wide range of theoretical arguments in support of the
relevance of linear wave properties even in a strongly turbulent
plasma, motivated by the mathematical properties of the nonlinear
equation of evolution for an incompressible MHD plasma.  We present an
analogy with the case of a critically damped simple harmonic
oscillator, and suggest that it is neither necessary nor expected that
one should see evidence of a linear dispersion relation ($\omega$
vs.~$k$) in measurements of turbulence. In addition, we present
argument that linear collisionless damping may persist even in the
strongly turbulent regime and the resulting plasma heating certainly can 
be spatially non-uniform, despite frequent claims to the contrary.

We review evidence in support of the quasilinear premise from
numerical simulations, including results supporting the applicability
of linear eigenfunctions and linear collisionless damping rates in the
turbulent plasma. Frequency diagnostics of plasma turbulence
simulations have yielded contradictory findings, but simulations
employing all three dimensions in space appear to largely find the
frequency of the turbulent dynamics consistent with the linear wave
frequencies. Given that the idea of critical balance in strong plasma
turbulence is essentially a quasilinear concept, evidence in support
of critical balance indirectly supports the quasilinear premise.
Simulations again yield contradictory results, but the line dividing
these conflicting results appear to coincide with the method used to
determine the direction of the magnetic field: studies using a
\emph{local} mean magnetic field are consistent with the predictions
of critical balance, while studies employing a global magnetic field
are not. 

Finally, we discuss observational evidence from turbulence in the
solar wind that supports the quasilinear premise, including
multi-spacecraft k-filtering analyses that find plasma-frame
frequencies of the turbulent fluctuations consistent with \Alfven and
kinetic \Alfven waves. Measurements of the the magnetic energy
spectrum as a function of the angle of the solar wind flow with
respect to the magnetic field also find support for critical balance
found when the magnetic field direction is determined locally.

The question of the validity of the quasilinear premise---that linear
wave properties are relevant to strong plasma turbulence---clearly
remains to be settled. On balance, however, we argue here that the
bulk of the evidence appears to support it as a valuable means of
modeling turbulence. The utility of the quasilinear premise for the
study of plasma turbulence, however, may also be judged \emph{a
  posteriori} by the insights gained from such an approach, and we
review a number of studies, including those using the novel
\emph{synthetic spacecraft data method}, that have succeeded in more
strongly constraining the fundamental nature of plasma turbulence.
The ultimate goal of turbulence models based on the quasilinear
premise is to develop the capability to predict the evolution of any
turbulent plasma system, including the spectrum of turbulent
fluctuations, their dissipation, and the resulting plasma heating.


%

\acknowledgments
This work was supported by NSF CAREER AGS-1054061 and NASA NNX10AC91G.




\clearpage

\end{document}